\documentclass[12pt]{article}
\usepackage{epsf}
\def\be{\begin{equation}}
\def\ee{\end{equation}}

\def\lsim{\raisebox{-0.6ex}{$\stackrel{\textstyle <}{\sim}$}}
\def\gsim{\raisebox{-0.6ex}{$\stackrel{\textstyle >}{\sim}$}}

\def\pp{\psi(2S)}
\def\jp{J/\psi}
\def\ypp{\Upsilon(2S)}
\def\yp1{\Upsilon(1S)}

\def\yp4{\Upsilon(4S)}

\begin{document}
\begin{titlepage}
\begin{center}
{\Large \bf William I. Fine Theoretical Physics Institute \\
University of Minnesota \\}
\end{center}
\vspace{0.2in}
\begin{flushright}
FTPI-MINN-11/01 \\
UMN-TH-2934/11 \\
January 2011 \\
\end{flushright}
\vspace{0.3in}
\begin{center}
{\Large \bf Hadronic transitions from $\Upsilon(4S)$ as a probe of four-quark admixture. 
\\}
\vspace{0.2in}
{\bf M.B. Voloshin  \\ }
William I. Fine Theoretical Physics Institute, University of
Minnesota,\\ Minneapolis, MN 55455, USA \\
and \\
Institute of Theoretical and Experimental Physics, Moscow, 117218, Russia
\\[0.2in]
\end{center}

\begin{abstract}

It is argued that the observed enhancement of the hadronic transition $\yp4 \to \Upsilon(1S) \, \eta$ relative to $\yp4 \to \Upsilon(1S) \, \pi \pi$ can be explained if the resonance $\yp4$ contains an admixture of a four-quark component, possibly due to the coupling to the meson channel $B \bar B$. A test of the discussed mechanism by searching for the transition $\yp4 \to \Upsilon(1S) \, \eta'(958)$ is suggested. Some properties of possible four-quark states related to the suggested one by the heavy quark spin symmetry are also discussed. 
\end{abstract}

\end{titlepage}

The resonance $\yp4$ is conventionally interpreted as an excited $4^3S_1$ state of bottomonium. However for a pure bottomonium it would be considerably difficult to understand the unusually large observed~\cite{babar} ratio of the rates for the hadronic transitions from $\yp4$ to $\Upsilon(1S)$: 
\be
\Gamma(\yp4 \to \Upsilon(1S) \, \eta)/\Gamma(\yp4 \to \Upsilon(1S) \, \pi^+ \pi^-)=2.41 \pm 0.40 \pm 0.12~.
\label{rep}
\ee
Indeed, the emission of the $\eta$ meson relative to that of a pion pair in transitions between states of a heavy quarkonium is suppressed by the flavor SU(3) symmetry and, for a $J^P=1^-$ quarkonium, by the heavy quark spin symmetry~\cite{mv07}. Such a suppression is prominently illustrated by the other observed similar ratio of the rates in bottomonium: $\Gamma(\ypp \to \Upsilon(1S) \, \eta)/\Gamma(\ypp \to \Upsilon(1S) \, \pi^+ \pi^-) \approx 10^{-3}$~\cite{cleo}. It can be also mentioned that in charmonium, where the heavy quark spin symmetry constraints are weaker, the suppression of the $\eta$ emission is also well known:  $\Gamma(\pp \to \jp \, \eta)/\Gamma(\pp \to \jp \, \pi^+ \pi^-) \approx 0.1$. The latter two processes involving lower lying states of the bottomonium and charmonium are reasonably well described by the QCD multipole expansion~\cite{gottfried,mv79}, which directly relates the decays with emission of $\eta$ to those with two pions~\cite{vz,mv07} and explicitly accounts for both types of suppression. A consideration based on the multipole expansion is not likely to be applicable to the transitions from $\yp4$ because of a larger spatial size of this bottomonium state and because of higher energy of the light hadrons in the transition to $\Upsilon(1S)$. However for a pure $b \bar b$ system, whether a $4^3S_1$ or with an admixture of $2^3D_1$~\cite{vbr} the emission of light mesons should still proceed through the gluonic field, and both the SU(3)$_{\rm fl}$ and the heavy quark spin suppression factors should be present. One can certainly notice that
in the transitions from $\ypp$ there is also a kinematical phase space suppression of the $\eta$ emission. Still, a scaling to the kinematics of the $\yp4 \to \Upsilon(1S)$ would be by far insufficient to enhance the ratio of the decay rates by a factor of thousands, necessary to explain the observed value (\ref{rep}). 

The purpose of the present paper is to point out that both suppression factors for the $\eta$ transitions disappear if the resonance $\yp4$ contains a four-quark admixture of an appropriate structure, and to discuss the possibilities of a further experimental investigation of the presence of such an admixture.

The amplitude of the $\eta$ transition between $J^P=1^-$ states has a unique structure in terms of the mixed product of the polarization amplitudes ${\vec \varepsilon}\, '$ and $\vec \varepsilon$ of the initial and the final states and the $\eta$ meson momentum $\vec p$:
\be
A(\yp4 \to \Upsilon(1S) \, \eta) = C \, ({\vec \varepsilon}\,' \times \vec \varepsilon^*) \cdot \vec p~,
\label{amp}
\ee
where $C$ is a constant, and the rest frame of the bottomonium is implied. For pure $^3S_1$ states of the quarkonium each polarization $\vec \varepsilon$ coincides with the total spin polarization amplitude of the heavy quark pair so that the amplitude (\ref{amp}) requires a spin-dependent interaction that would rotate the total spin. Therefore one concludes that the constant $C$ should be proportional to the inverse of the heavy quark mass, $m_b^{-1}$. It is clear that an admixture of a $^3D_1$ wave function does not invalidate this conclusion, since in the absence of a spin-dependent interaction in the decay, the spin of the heavy quark pair would go through, and what would be left is an emission of the $\eta$ meson by a transition from the $J^P=2^+$ $D$ wave orbital spinless state to an $J^P=0^+$ $S$ wave orbital state and the $J^P=0^-$ $\eta$ meson, which is obviously forbidden.

It is therefore compelling to conclude that the $\eta$ transition from $\yp4$ to $\Upsilon(1S)$ proceeds due to an admixture of a state containing light quarks in addition to the $b \bar b$ pair in the $\yp4$. Some amount of such an admixture with non-strange light quark states should be expected on general grounds due to the coupling of $\yp4$ to the open heavy flavor channel $B \bar B$. In this respect the similarity between the $\yp4$ and $\psi(3770)$ may extend beyond both resonances being similarly located at the corresponding heavy flavor threshold and extend to a similar four-quark content that can be also viewed as being due to a re-annihilation~\cite{rosner} of the heavy mesons. In the case of $\psi(3770)$ there is also an observed~\cite{cleo05} enhancement of the $\eta$ transition $\psi(3770) \to \eta \, \jp$, in line with the expectations based on estimates~\cite{mv05} of the four-quark admixture effects in the $\psi(3770)$. Unlike in the charmonium, where the effect of the four-quark component is less certain due to the heavy quark spin selection rule being not as strong, the relative enhancement of the $\eta$ transition from the $\yp4$ may enable a more thorough analysis of the four-quark part of the wave function that is responsible for the transition.

Indeed, in the limit of an exact heavy quark spin conservation the amplitude with the structure (\ref{amp})  can only arise if the wave function of $\yp4$ contains a part that can be represented as a composition of a $J^P=1^-$ heavy quark pair and a $J^P=1^+$ state of the light matter: $(1^-)_H \otimes (1^+)_L$.  In other words, the $b \bar b$ pair is in a $^3S_1$ state and in addition there are light degrees of freedom with the quantum numbers $J^P=1^+$ so that the two subsystems combine in a $J^{PC}=1^{--}$ overall system. Using the notation $\chi_i$ for the spin variables of the $^3S_1$ heavy pair and $\ell_j$ for the polarization wave function of the light $1^+$ component the polarization amplitude $\psi_k$ of the four-quark admixture in $\yp4$ can be written as
\be
\psi_k = {1 \over \sqrt{2}} \,\epsilon_{ijk} \, \chi_i \, \ell_j~.
\label{pce}
\ee
It can be noted that neither the heavy nor the light part of the four-quark component needs to be assumed as being colorless, since an exchange of color by the gluon interaction is not suppressed in such system. Also the overall quantum numbers $J^P=1^+$ include both the internal spin-parity part of the light quark-antiquark pair as well as the possible orbital momentum of the pair as a whole in the rest frame of the heavy quarks. The mixing between the possible states resulting in the overall $1^+$ quantum numbers of the light degrees of freedom is not suppressed, since neither spin nor orbital excitation of the light quarks requires any suppression. 

Clearly, for a system with the structure as in Eq.(\ref{pce}) the amplitude (\ref{amp}) is generated by $\psi_k$ being proportional to the initial state polarization $\varepsilon'_k$, while the heavy quark spin part $\chi_i$ materializes as the polarization of the final $\Upsilon(1S)$: $\varepsilon^*_i$, and, finally, the light subsystem produces the $\eta$ meson as $\ell_j \to p_j$. No spin dependent interaction for the heavy quark is required and the flavor symmetry suppression is also lifted, since the light quarks in the discussed admixture are most likely to be only the non strange ones. It can as well be noticed that in the leading order in the heavy quark symmetry the decay  $\yp4 \to \Upsilon(1S) \, \pi \pi$ is not affected by the presence of the discussed four-quark component, and can proceed due to a more conventional gluonic mechanism.

The assumed structure of the four-quark part in the $\yp4$ resonance suggests that there should be yet unobserved transitions from the $\yp4$ due to decay of the light component, which also would not require a spin or orbital excitation of the heavy quarks. Namely, the transition $\yp4 \to \Upsilon(1S) \, \eta'(958)$ is kinematically allowed and should be enhanced similarly to $\yp4 \to  \Upsilon(1S) \, \eta $. The relative strength of the $\eta'$ and $\eta$ transitions is also a probe of the light-flavor composition of the discussed four-quark component. Indeed, if, as one can reasonably assume, the admixture of the light degrees of freedom arises from the coupling of the $\yp4$ to $B \bar B$ meson pairs, only non strange quarks are present in the light component, so that the $SU(3)_{\rm fl}$ symmetry is badly broken~\footnote{Any significant violation of the isotopic symmetry is not expected, in contrast to the case of the $\psi(3770)$ where the large isotopic mass difference for the $D$ mesons is likely essential~\cite{mv05}.}, and the ratio of the transition rates can be estimated from the relative $u \bar u + d \bar d$ content of the $\eta$ and $\eta'$ mesons. Using the standard $SU(3)_{\rm fl}$ expression in terms of the $\eta-\eta'$ mixing angle $\theta_{\eta \eta'} \approx 0.2$:
\begin{eqnarray}
&& \eta= \cos \theta_{\eta \eta'} \, {u \bar u + d \bar d - 2 \, s \bar s \over \sqrt{6}}+ \sin \theta_{\eta \eta'} \, {u \bar u + d \bar d + \, s \bar s \over \sqrt{3}}\,, \nonumber \\
&&\eta'= -\sin \theta_{\eta \eta'} \, {u \bar u + d \bar d - 2 \, s \bar s \over \sqrt{6}} + \cos \theta_{\eta \eta'} \, {u \bar u + d \bar d + \, s \bar s \over \sqrt{3}}\, ,
\end{eqnarray}
one can readily estimate the expected ratio of the transition rates
\be
{ \Gamma(\yp4 \to \Upsilon(1S) \, \eta') \over \Gamma(\yp4 \to \Upsilon(1S) \, \eta )} =  \left ( {\sqrt{2}- \tan \theta_{\eta \eta'} \over 1+ \sqrt{2} \, \tan \theta_{\eta \eta'}} \right )^2 \, {p_{\eta'}^3 \over p_\eta^3} \, \left | {F(p_{\eta'}) \over F(p_\eta)} \right |^2 \approx 0.2 \, \left | {F(p_{\eta'}) \over F(p_\eta)} \right |^2~,
\label{rat}
\ee
where $F$ is the form factor. Given that the inelasticity in the $0^-$ channel is not strong below $1\,$GeV, the deviation of this form factor from one at the momentum of $\eta$ ($p_\eta=$924\,MeV) should not be large. The form factor generally suppresses the emission of higher momentum particles, so that one can estimate
\be
0.2 \, \lsim \, { \Gamma(\yp4 \to \Upsilon(1S) \, \eta') \over \Gamma(\yp4 \to \Upsilon(1S) \, \eta )} \, \lsim \, 0.6~,
\label{rate}
\ee
where the upper estimate corresponds to a form factor suppression of the $\eta$ emission by a factor of three in the rate.

One can notice that for the standard mechanism of hadronic transitions in heavy quarkonium through the emission of light mesons by the gluonic field, an estimate of the ratio (\ref{rat}) would be based on the $SU(3)_{\rm fl}$ symmetry and would contain instead of the factor $(\sqrt{2}- \tan \theta_{\eta \eta'})^2/(1+ \sqrt{2} \, \tan \theta_{\eta \eta'})^2 \approx 1$ the factor $\theta_{\eta \eta'}^{-2} \sim 25$.  Thus in this case the numerical estimate of the ratio would be by a factor of approximately 25 higher than in Eq.(\ref{rate}). The well known examples of such a significant dominance of the $\eta'$ emission by the gluonic mechanism over that of $\eta$ are provided~\cite{pdg} by the radiative decays of charmonium: $\Gamma(\jp \to \gamma \, \eta')/ \Gamma(\jp \to \gamma \, \eta) \approx 4.8$, $\Gamma(\pp \to \gamma \, \eta')/ \Gamma(\pp \to \gamma \, \eta) \, \gsim \, 50$. 

Another possible option for the light meson final state produced by the discussed light component in the $\yp4$ is the transitions $\yp4 \to \Upsilon(1S) \, \eta \, \pi \, \pi$ with the light-meson system having quantum numbers $0^-$ or $1^+$. However at this point I am not aware of any way to estimate the possible rate of such processes.

In the suggested four quark structure of the type $(1^-)_H \otimes (1^+)_L$ the angular momenta of the heavy and light components combine into the overall quantum numbers $1^{--}$ inside the $\yp4$ resonance. This suggestion naturally invites a question about similar structures that are related to the discussed one in the first order in the interaction depending on the spin of the $b$ quark, which interaction is suppressed  as $m_b^{-1}$. Namely, the two other states of the type $(1^-)_H \otimes (1^+)_L$ with the overall quantum numbers $J^{PC}=0^{--}$ and $2^{--}$ as well as that of the type $(0^-)_H \otimes (1^{+})_L$ with $J^{PC}= 1^{-+}$ differ from the considered one by a re-orientation of the spin of the heavy quarks relative to the overall spin of the light component. The splitting in energy between such states should be of order $\Lambda_{QCD}^2/m_b$. If they existed, the first two states would have transitions to $\Upsilon(1S) \, \eta$ and $\Upsilon(1S) \, \eta'$, while the third one would be able to decay into $\eta_b \, \eta$ and $\eta_b \, \eta'$. It can be also noticed that unlike in a pure heavy quarkonium where a flip of the heavy quark pair from the spin triplet to the spin singlet state is of second order in the heavy quark spin interaction (e.g. $\Upsilon - \eta_b$ splitting), in a four quark system such a flip is only of the first order in $m_b^{-1}$, due to the interaction with the light component~\footnote{In other words, it is the re-orientation of heavy quark spin relative to the angular momentum of the light component which produces the dominant effect, rather than the change in the spin-spin interaction between the heavy quarks.}. All such additional states have exotic quantum numbers and cannot be realized in a $B \bar B$ meson pair. If it is indeed the coupling to the $B \bar B$ channel that results in the discussed four-quark admixture in $\yp4$ then such states should not be present in the spectrum near or below the $B \bar B$ threshold, although they can arise near the higher meson thresholds, e.g. $B^* \bar B$, whose energy separation from $\yp4$ is also of order $\Lambda_{QCD}^2/m_b$. Naturally, these exotic resonances are not accessible in the $e^+e^-$ experiments, but can be visible if the $b \bar b$ pairs are produced by other sources such as the high-energy $pp$ collisions studied by the  LHCb experiment. These extra states could in principle be observed either in the $B^* \bar B$ invariant mass spectra, if they are above the meson threshold, or if they are below the threshold, by their hadronic transitions to $\Upsilon(1S) \, \eta (\eta')$ and $\eta_b \, \eta (\eta')$ respectively. If, on the other hand, the existence of a pure quarkonium state is critical for such a four-quark system to arise as an admixture, such extra states would not be possible due to their exotic quantum numbers, and the four-quark system within the $\yp4$ would be the only one in existence.

In summary. It is argued here that the observed large rate of the transition  $\yp4 \to \Upsilon(1S) \, \eta$ relative to that of $\yp4 \to \Upsilon(1S) \, \pi^+ \pi^-$ is at a considerable variance from the expected for pure $b \bar b$ states suppression of the $\eta$ emission by heavy quark spin symmetry and flavor $SU(3)_{\rm fl}$. Both suppression factors do not apply if the resonance $\yp4$ contains an admixture of a four-quark state with the spin structure $(1^-)_H \otimes (1^+)_L$ which can arise, e.g. due to the coupling to the open-flavor meson channel $B \bar B$. If this is indeed the mechanism responsible for the $\eta$ emission, one should expect a similar transition with the emission of the $\eta'$ meson to have a rate estimated in Eq.(\ref{rate}) as being somewhat below the rate of the decay $\yp4 \to \Upsilon(1S) \, \eta$, while for a standard emission of the light mesons by gluonic field the relative probability of the $\eta'$ transition can be expected to be significantly larger than that for $\eta$. It appears to be quite realistic that the discussed decay $\yp4 \to \Upsilon(1S) \, \eta'(958)$ can be searched for by an analysis of the already accumulated data at the B factories.

This work is supported in part by the DOE grant DE-FG02-94ER40823.

\end{document}